\documentclass[
3p,
final,
nopreprintline,
times,number,sort&compress,lefttitle]{elsarticle}
\usepackage{graphicx}
\usepackage[english]{babel}

\usepackage{fancyhdr}
\newlength{\figsize}
\begin{document}

\title{Possibility of negative refraction for visible light in disordered all-dielectric resonant high index metasurfaces\tnoteref{alink}}
\tnotetext[alink]{https://doi.org/10.1016/j.ijleo.2019.163739}
\author{Andrey V. Panov}

\ead{andrej.panov@gmail.com}
\address{
Institute for Automation and Control Processes,
Far East Branch of Russian Academy of Sciences,
5, Radio st., Vladivostok, 690041, Russia}

\begin{abstract}

Effective refractive index of disordered all-dielectric metasurfaces consisting of gallium phosphide (GaP) spheres is studied by means of three-dimensional finite-difference time-domain (FDTD) simulations at the wavelength of 532~nm. 
It is shown that a mixture of the high index nanoparticles with sizes close to the first magnetic and electric resonances randomly dispersed on metasurface may possess negative refraction. 
The dependence of the metasurface effective refractive index on the nanoparticle concentration and size is constructed.
The feasibility of negative refraction at large concentrations of the high index resonant spheres is demonstrated.
The negative effective refractive index is exhibited only for a monolayer of GaP spheres.

\end{abstract}

\maketitle

In recent years, the research in all-dielectric resonant nanophotonics is a rapidly developing. 
The all-dielectric photonic nanostructures offer a significant advantage over plasmonic structures due to low losses. 
The all-dielectric metamaterials with the negative effective index of the refraction were proposed in the early 2000s mainly for the terahertz range.
Initially, the all-dielectric metamaterials were predicted theoretically \cite{Holloway03,Vendik04,Jylha06} with the use of Lewin's \cite{Lewin47} or similar expressions for the effective properties of an ordered array of spherical particles embedded in a matrix.
Subsequently, the all-dielectric nanostructures with negative refraction were realized experimentally \cite{Peng07,Lai09,Wang12}.
In the past few years, the dielectric metasurfaces with negative refraction have been experimentally implemented. 
These metasurfaces are realized as regular lattices of the scattering elements with subwavelength sizes. 
Currently, the shape and size of the high index particles can be precisely controlled \cite{Verre18}. 
However, precious positioning of these elements is not an easy task. 
Whereas, manufacturing the metasurfaces of randomly arranged scatterers is a much simpler process. Currently, the disordered metasurfaces are shown to enhance the lens resolution at visible wavelengths because of wavefront shaping \cite{vanPutten11,Jang18}. The modeling of random metasurfaces of GaP spheres near Mie resonances exhibit enormous optical Kerr effect due to field concentration \cite{Panov18a,
Panov19}.
The possibility of negative refraction in disordered all-dielectric metamaterials was theoretically studied in Ref.~\cite{Slovick17}. 
The author concluded that the negative refraction can be observed at large concentrations of scatterers, whereas the acceptable transmission of the disordered metamaterial should be limited to a monolayer.
Nevertheless, the presented in work \cite{Slovick17} analysis is three-dimensional.
Moreover, the results of FDTD simulations of light transmission through thick particulate samples show significant discrepancies between refractive index calculated in Ref.~\cite{Panov19} and the result of the effective medium theory \cite{Slovick17}. 
It may be due to an assumption of zero scattering from the composite material while this is questionable for the resonant particles.
Theory of Ref.~\cite{Slovick17} was tested only for inclusions with sizes below the resonances.
It is noteworthy that the use of the Lewin's homogenization formula for such composite materials was argued in Ref.~\cite{Mackay08} since wavelength inside high index inclusions is comparable with their size.
In this work, the possibility of realization of the disordered all-dielectric resonant high index metasurfaces with the negative refraction of visible light is demonstrated by numeric modeling.

The refractive index of metasurfaces was retrieved with a method proposed in Ref.~\cite{Panov18}. 
This method is based on calculating the phase change of the Gaussian beam transmitted through the studied sample. 
The sample introducing the phase change can be treated as a continuous material with the effective refractive index.
In this work, the effective refractive index represents the ensemble-averaged optical characteristics of the sample. 
Sometimes, it is called the equivalent refractive index and it should not be related to the effective medium approximations.
Initially, the technique is elaborated for obtaining the nonlinear refractive index but it can be utilized for the evaluation of the real part of the linear index of refraction $n$. 
Since the phase in Ref.~\cite{Panov18} is calculated at the axis of the Gaussian beam at large distances from the sample the most scattered light does not affect the calculated value of $n$.
At present, the effective permittivity and permeability are usually derived from the complex reflection and transmission coefficients ($S$ parameters) \cite{Smith02}.
For resonant samples, the planar wavefronts used in \cite{Smith02} are significantly distorted due to substantial scattering, so the $S$ parameters technique shows large deviations.
To obtain the similar accuracy of $n$, the method based on the calculation of the $S$ parameters requires approximately twice the computational domain compared to that used in Ref.~\cite{Panov18}. 
Moreover, the error of restoration of the effective extinction coefficient of GaP nanocomposite using $S$ parameters always exceeds the estimated value owing to low GaP extinction coefficient in the visible range. 
Hence, the method established in Ref.~\cite{Smith02} gives no benefits over the computation of the transmitted Gaussian beam phase change.
The extinction coefficient $\kappa$ can be roughly estimated using transmittance \cite{Sato67,Sopori98}
\[
T=\exp(-4\pi \kappa L/\lambda)
\]
where $L$ is the thickness of the specimen, $\lambda$ is the wavelength in free space. 
In fact, this estimate gives the upper limit of the extinction coefficient since the reflectance is disregarded.

The size of the computational domain for the three-dimensional FDTD simulations was $4\times 4\times 30$~$\mu$m with the space resolution of 5~nm. 
The radius $w_0$ at the Gaussian beam waist was $1.1$~$\mu$m. 
These parameters made it possible to calculate with a good degree of accuracy the refractive index of the high index resonant metasurfaces. 

The gallium phosphide (GaP) was selected for modeling high index inclusions in the present study as this material has a moderately low extinction coefficient in the visible range. 
According to Ref.~\cite{Aspnes83}, at  wavelength $\lambda=532$~nm the complex refractive index of GaP is $n_\mathrm{in}+i\kappa_\mathrm{in}=3.49+0.0026i$.
The studied samples represent the separated GaP spheres randomly arranged on plane and surrounded by free space. 
The procedure of generating the disoredered arrangement of the spheres continued until the desired number of the particles is achieved providing the exclusion of the possibility of their overlaps.
The simulated Gaussian beam falls at normal incidence on the metasurface.

Subwavelength dielectric spheres with the high refractive index at Mie resonances support modes with the self-maintaining oscillations of the electric and magnetic fields \cite{Debye1909,Hulst81}. 
The first resonance with the lowest size of dielectric particles with relative permittivity $\varepsilon>0$ is a magnetic dipole resonance which is most pronounced.
In accordance with Refs.~\cite{Debye1909,Hulst81}, the magnetic and electric resonances in high-index spheres are observed when
\[
\frac{2\pi r n_\mathrm{in}}{\lambda}=c_{j-1},\quad \frac{2\pi r n_\mathrm{in}}{\lambda}=c_j \left( 1 - \frac{1}{n_\mathrm{in}^2 j}\right) ,
\]
where $n_\mathrm{in}$ is the refractive index of inclusions, $r$ is their radius, $j=1,2,\ldots$, $c_0=\pi$, $c_1=4.493$.
The magnetic dipole and quadrupole Mie resonances for the standalone GaP spheres at the wavelength of $532$~nm exist at the radii of $r=76$ and 109~nm, the lowest electric resonances are expected at $r=100$~nm and $134$~nm.

\begin{figure}
{\centering
\includegraphics[width=8cm]{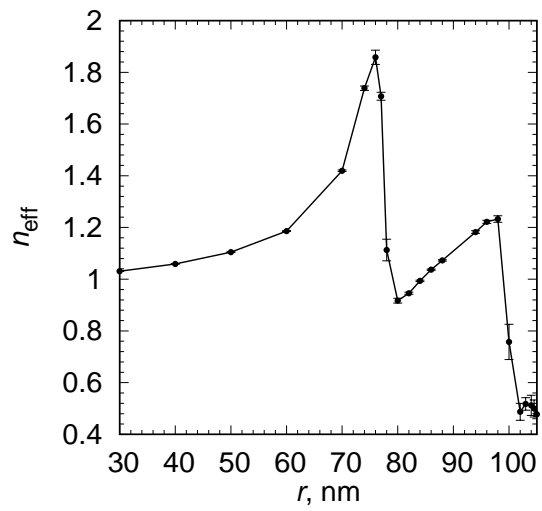}
\par} 
\caption{\label{n0_r_GaP_loss_180_randsurf}Dependence of the effective refractive index $n_{\mathrm{eff}}$ of the metasurface on the sphere radius~$r$ for 180 identical particles randomly positioned on $4\times 4$~$\mu$m area. The error bars show the standard deviations of $n_{\mathrm{eff}}$.} 
\end{figure} 

Fig.~\ref{n0_r_GaP_loss_180_randsurf} depicts the  dependence of the refractive index of a disordered metasurface with GaP spheres equal in size. 
As can be seen, this dependence is non-monotonic since the first magnetic and electric Mie resonances result in dips of the curve. 
The sizes for these resonances are approximately equal to those predicted by the Mie theory. 
Alternatively, in a periodic lattice of the particles with the large volume fraction of inclusions, the resonance sizes may be changed substantially (see, e.g. \cite{Silveirinha11}). 
The main reason for this is interparticle interactions at large concentrations of the inclusions.
There exist ranges of particle sizes just above the magnetic resonances with negative effective magnetic permeability $\mu_\mathrm{eff}$. 
In a similar manner, effective electric permittivity $\varepsilon_\mathrm{eff}$ of the spheres with diameters slightly higher than the size of the electric dipole resonance is below zero. 
The all-dielectric metamaterials with negative refraction are designed by combining both resonances in one medium. One possible way to achieve this goal is overlapping the magnetic and electric Mie resonances for a lattice of identical particles at certain volume fractions and frequencies \cite{Huang18}. 
The disordered metasurface of GaP spheres does not show the ability to combine two resonances for the particles of the same size. 
Nevertheless, the mixture  of GaP spheres with two radii might exhibit the negative index of refraction.

\begin{figure}
{\centering
\includegraphics[width=8cm]{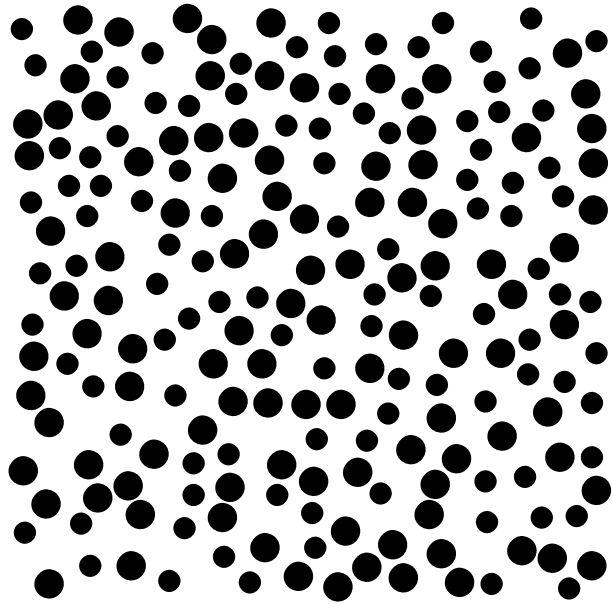}
\par} 
\caption{\label{rand_bidisperse}
Disordered bidisperse metasurface.} 
\end{figure} 

\begin{figure}
{\centering
\includegraphics[width=8cm]{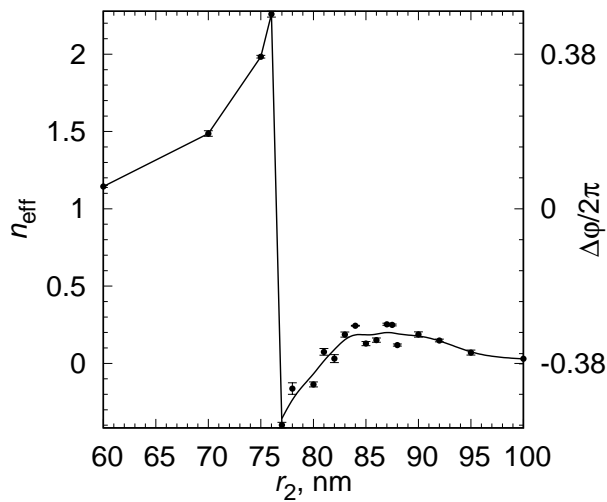}
\par} 
\caption{\label{n0_rx-101_GaP_loss_randsurf}Effective refractive index $n_{\mathrm{eff}}$ of the disordered metasurface consisting of two types of GaP spheres: with the fixed radius of $r_1=101$~nm and the varying second radius $r_2$ as abscissa. The second $y$-axis displays the phase change $\Delta \varphi$ introduced by the metasurface. The number of particles is 264 on $4\times 4$~$\mu$m area.} 
\end{figure} 

Further, the random metasurface consisting of equal numbers of two kinds of GaP spheres was modeled: the first ones with the radius of 101~nm (near the first electric resonance) and the second particles with varying radius in the vicinity of the magnetic dipole resonance. 
The typical random bidisperse arrangement of the spheres is illustrated by Fig.~\ref{rand_bidisperse}.
The effective refractive index of the metasurface as a function of the second radius is shown in Fig.~\ref{n0_rx-101_GaP_loss_randsurf}.
The number of nanoparticles was 264 on $4\times 4$~$\mu$m area which is close to maximum packing of randomly positioned spheres.
It should be emphasized that there is a range of radii of the spheres of the second type above the magnetic dipole resonance (77--80~nm) with negative $n_{\mathrm{eff}}$.
Thus, the mixture of the high index spheres with sizes close to the first electric and magnetic Mie resonances may possess $n_{\mathrm{eff}}<0$.

Instead of using $n_\mathrm{eff}$, the phase shift of the transmitted light can be utilized (see, e.q. Ref.~\cite{Yu15}). 
The disordered metasurface studied in this work show the ability to change this phase shift in the range of $2\pi$. 
In contrast to Ref.~\cite{Yu15}, the phase shift exhibits discontinuity at the resonances. 
The magnetic and electric dipole resonances in short cylinders used in Ref.~\cite{Yu15} are less pronounced in comparison with spheres.

\begin{figure}
{\centering
\begin{minipage}[b]{\figsize}
\includegraphics[width=\figsize]{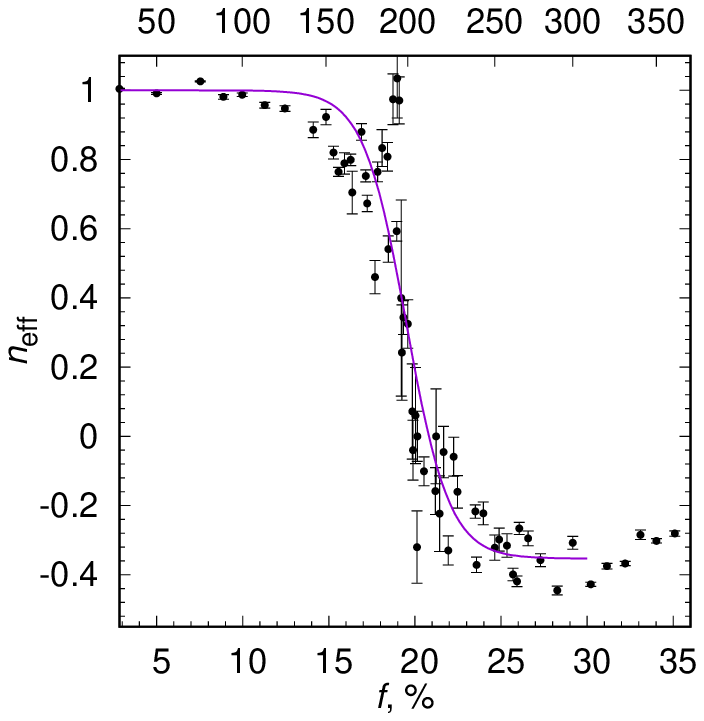}
\end{minipage}\hfill
\begin{minipage}[b]{\figsize}
\includegraphics[width=\figsize]{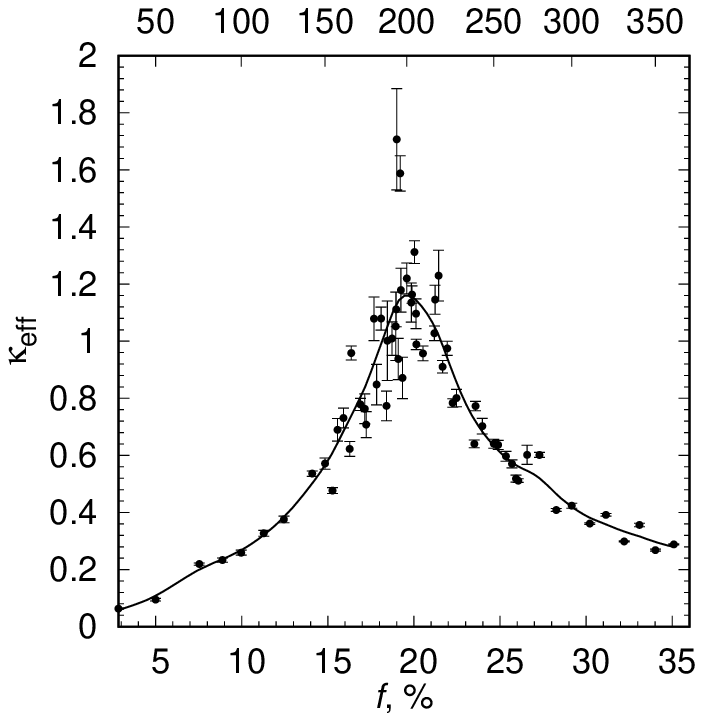}
\end{minipage}\par} 
\caption{\label{n0_conc_r77-101_GaP_loss_surf_c}Effective refractive index $n_{\mathrm{eff}}$ and extinction coefficient $\kappa_{\mathrm{eff}}$ of the random metasurface consisting of the equal numbers of GaP spheres with the radii of 77 and 101~nm versus volume fraction of the particles $f$. The upper abscissa axes display the net number of the particles on $4\times 4$~$\mu$m area. The error bars show the standard deviations of $n_{\mathrm{eff}}$.} 
\end{figure}

\begin{table}
{
\centering 
\begin{tabular}{llc}
\hline
$f$, \% &$a$,~nm&$n_{\mathrm{eff}}$\\
\hline
18.1& 600& $\hphantom{-}0.86 \pm 0.04$\\
21.9& 540& $\hphantom{-}0.53 \pm 0.11$\\
27.5& 480& $-0.11 \pm 0.01$\\
32.3& 440& $-0.19 \pm 0.01$\\
38.8& 400& $-0.13 \pm 0.01$\\
\hline
\end{tabular}
\par}
\caption{\label{n0_conc_r77-101_GaP_loss_squarelatt}
Effective refractive indexes of square lattices of GaP spheres with the radii of 77 and 101~nm. 
Here $a$ is a lattice constant, $f$ is volume fraction of the particles.}
\end{table}

Fig.~\ref{n0_conc_r77-101_GaP_loss_surf_c} illustrates effective refractive index $n_{\mathrm{eff}}$ and extinction coefficient $\kappa_{\mathrm{eff}}$  of the disordered metasurface comprising the equal numbers of GaP spheres of two radii (77 and 101~nm) as functions of volume fraction of the particles $f$. 
Herein, $f$ is defined as ratio of the net volume of the particles to the volume of the smallest cuboid containing these inclusions.
Here and in the following, each point on graphs corresponds to a separate bidisperse random arrangement of particles.
At low concentrations of GaP spheres, $n_{\mathrm{eff}}$  tends to unity. 
Then with growth in $f$, the effective index gradually decreases and becomes negative approximately at $f=20\%$. 
Further, $n_{\mathrm{eff}}$ is negative reaching the lowest value about $-0.40$. 
This behavior is similar to a phase transition from random scattering on rare particles to the metamaterial based on the Mie resonances.
The large standard deviations for values of $n_{\mathrm{eff}}$ close to zero should be noted.
At volume fraction above 30\%, the GaP spheres become touch each other thus the conditions of the resonances do not fulfill.
Thus, $n_{\mathrm{eff}}$ has tendency to enlarge in this range.
The dispersion of the points in Fig.~\ref{n0_conc_r77-101_GaP_loss_surf_c} arises from configuration specific effects.
Before, Felbacq and Bouchitt\'e \cite{Felbacq05} demonstrated the possibility of the negative refraction in two-dimensional random photonic crystals consisting of square rods with permittivity $\varepsilon=200+5i$, but such material is impossible  for visible wavelengths. 
As shown in Ref.~\cite{Felbacq05}, the disorder on the position of the rods does not suppress the negative refraction.
To compare with results presented in Fig.~\ref{n0_conc_r77-101_GaP_loss_surf_c}, the computed magnitudes of $n_{\mathrm{eff}}$ for square lattices consisting of two sublattices with GaP spheres with the radii of 77 and 101~nm are given in Tab.~\ref{n0_conc_r77-101_GaP_loss_squarelatt}. 
As is obvious, the values of $n_{\mathrm{eff}}$ for square lattices are higher than that of the random metasurface. 
In other words, the disordered metasurfaces allow one to obtain the lower values of negative refraction with respect to lattices.

The estimates of $\kappa_{\mathrm{eff}}$ (Fig.~\ref{n0_conc_r77-101_GaP_loss_surf_c}) show moderate values for volume fractions with negative $n_{\mathrm{eff}}$.
These magnitudes do not exceed those observed for plasmonic negative index metamaterials.
It should be stressed that significant loss is inherent to the negative refraction \cite{Stockman07}.

\begin{figure}
{\centering
\includegraphics[width=\figsize]{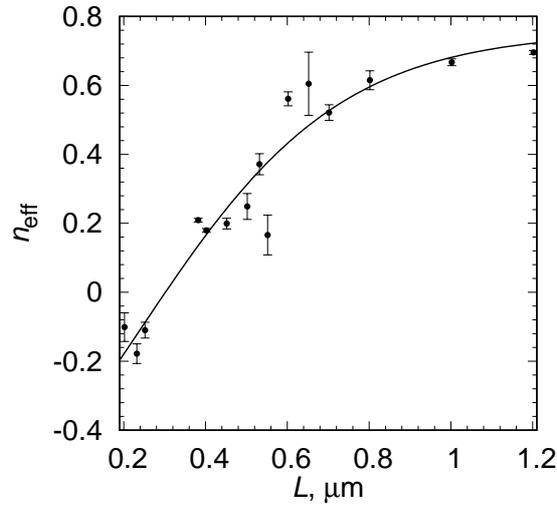}
\par} 
\caption{\label{n0_ML_r77-101_GaP_loss_surf_f20.5}Effective refractive index $n_{\mathrm{eff}}$ of the random nanocomposite consisting of GaP spheres with the radii of 77 and 101~nm as a function of its thickness $L$ at fixed value of volume fraction of the particles $f=20.5\:\%$. The error bars show the standard deviations of $n_{\mathrm{eff}}$. The thickness of the monolayer is 0.202~$\mu$m.} 
\end{figure} 

Fig.~\ref{n0_ML_r77-101_GaP_loss_surf_f20.5} depicts the dependence of the effective refractive index of the disordered nanocomposite on its thickness when the volume fraction of the particles is fixed ($f=20.5\:\%$). There is a range of thicknesses (250--380~nm) when it is not possible to realize the composite with this volume fraction.
This range becomes wider for higher volume fractions.
The effective refractive index showed a tendency to increase with the thickness of the nanocomposite. 
The values of $n_{\mathrm{eff}}$ saturate with increasing thickness.
The negative values of $n_{\mathrm{eff}}$ are exhibited only when the thickness of the nanocomposite is close to one monolayer (202~nm).  

\begin{figure}
{\centering
\includegraphics[width=\figsize]{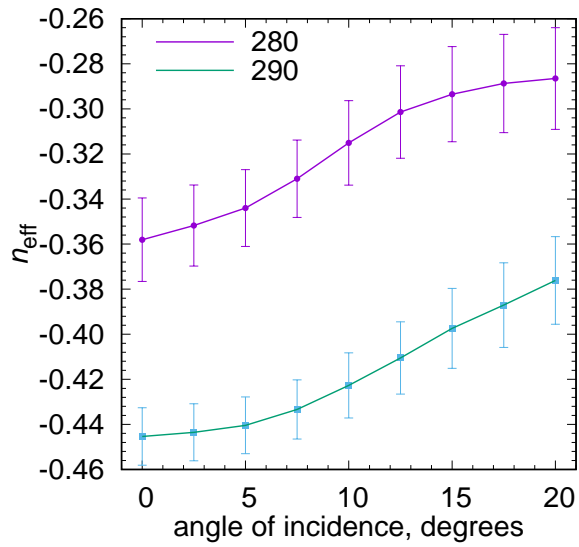}
\par} 
\caption{\label{n0_deg_r77-101_GaP_loss_280_randsurf}Effective refractive index $n_{\mathrm{eff}}$ of the random nanocomposite consisting of GaP spheres with the radii of 77 and 101~nm as a function of angle of incidence of the beam. The error bars show the standard deviations of $n_{\mathrm{eff}}$. The net number of the particles on $4\times 4$~$\mu$m area is shown in the graph.} 
\end{figure} 

The method based on the calculation of the phased change of the transmitted beam is designed for normal incidence. But it can be applied for small angles of incidence with reasonable accuracy. The dependence of $n_{\mathrm{eff}}$  on the angle of incidence is illustrated by Fig.~\ref{n0_deg_r77-101_GaP_loss_280_randsurf}. There is no drastic change of $n_{\mathrm{eff}}$ at the small angles.
The decrease of $|n_{\mathrm{eff}}|$ mostly arises from enlarging the optical path in the inclined metasurface.

In particular, by varying the concentration of the nanoparticles on the metasurface, it is possible to design the metasurfaces with different values of the refractive index. 
For example, all-dielectric zero index metamaterial can be realized. 
Currently, zero index metamaterials have attracted much interest of researchers \cite{Moitra2013,Liberal17}.
The use of disordered metasurfaces is expected to simplify  the fabrication process of metamaterials with tailored optical properties.
It is essential that at other wavelengths the sizes of the resonances as well as the values of $n_\mathrm{in}+i\kappa_\mathrm{in}$ are different so the analysis has to be done again.

The negative refraction by the metasurface may be used for constructing a perfect lens as supposed in Ref.~\cite{Pendry00}. The all-dielectric zero index metamaterial is expected to have enhanced nonlinear properties which were observed for thin films with epsilon-near-zero conditions \cite{Caspani16,Alam16}.

In conclusion, the effective refractive index of random GaP metasurface is numerically studied as a function of the high index particle size and concentration. 
It is shown that the disordered mixture of the spheres with sizes close to the first magnetic and electric resonances has negative index of refraction in the visible range at large volume fractions of the nanoparticles. 
Just the monolayer metasurface is demonstrated to possess the negative refractive index.
The random high index metasurfaces allows one to achieve lower values of the negative refractive index as compared to that of the ordered square lattices.

The results were obtained with the use of IACP FEB RAS Shared Resource Center ``Far Eastern Computing Resource'' equipment (https://www.cc.dvo.ru). 

\bibliography{nlphase}
\end{document}